\documentclass[letterpaper, 10 pt, conference]{ieeeconf}  

\IEEEoverridecommandlockouts                              
\usepackage{amsmath} 

\usepackage{xfrac}
\usepackage{bigints}
\usepackage{amssymb}  
\usepackage{mathtools}
\usepackage{graphicx}
	\graphicspath{ {./Pics/} }
\usepackage{float}
\usepackage[export]{adjustbox}

\newcommand{\R}{\mathbb{R}}

\title{\LARGE \bf Contingency Model Predictive Control for Automated Vehicles}

\author{John P. Alsterda$^{1,2}$, Matthew Brown$^{1}$ and J. Christian Gerdes$^{1}$
\thanks{$^{1}$Department of Mechanical Engineering, Stanford University, Stanford, CA 94305, USA
        {\tt\small}}%
\thanks{$^{2}$Corresponding author: {\tt\small alsterda@stanford.edu}}}

\begin{document}

\maketitle
\thispagestyle{empty}
\pagestyle{empty}

\begin{abstract}

We present Contingency Model Predictive Control (CMPC), a novel and implementable control framework which tracks a desired path while simultaneously maintaining a contingency plan -- an alternate trajectory to avert an identified potential emergency. In this way, CMPC anticipates events that \textit{might} take place, instead of reacting when emergencies occur. We accomplish this by adding an additional prediction horizon in parallel to the classical receding MPC horizon. The contingency horizon is constrained to maintain a feasible avoidance solution; as such, CMPC is \textit{selectively} robust to this emergency while tracking the desired path as closely as possible. After defining the framework mathematically, we demonstrate its effectiveness experimentally by comparing its performance to a state-of-the-art deterministic MPC. The controllers drive an automated research platform through a left-hand turn which may be covered by ice. Contingency MPC prepares for the potential loss of friction by purposefully and intuitively deviating from the prescribed path to approach the turn more conservatively; this deviation significantly mitigates the consequence of encountering ice.

\end{abstract}

\section{INTRODUCTION} 

In a future when automated vehicles are ubiquitous, we will expect them to handle the surprises and challenges that humans tackle routinely. They must avoid children who may run into the street, or maneuver safely when encountering an icy surface. In control engineering, this concept is known as robustness: a controller's ability to safely operate when subject to unexpected variation in system or environmental parameters. In the field of Model Predictive Control (MPC), work to this end falls largely into two broad categories: Robust MPC (RMPC) and Stochastic MPC (SMPC). 

In two reviews, Mayne described some RMPC strategies, their limitations, and direction for future development \cite{Mayne2014ModelPromise}\cite{Mayne2016RobustDirection}. In these robust algorithms, uncertain parameters' values are first confined to a set between upper and lower bounds. A control trajectory is then calculated which satisfies system constraints for \textit{all} possible combinations and realizations of these parameters. Thus RMPC is sometimes dubbed `worst case' MPC, prepared for the most extreme coincidences of bad luck -- however unlikely.

RMPC is often cast as a min-max problem, solving for commands that minimize the maximum possible cost \cite{Campo1987ROBUSTCONTROL}. The problem is in general nonlinear and non-convex, but Sartipizadeh \textit{et~al.} applied this technique on an approximate convex hull of uncertainty to achieve computational feasibility \cite{Sartipizadeh2016UncertaintyMethod}. D.~He \textit{et~al.} also ensured the tractability of their RMPC work by employing `quasi-min-max' MPC \cite{He2014Quasi-min-maxStability}.

\begin{figure}[t]
	\centering
    \includegraphics[scale=.25]{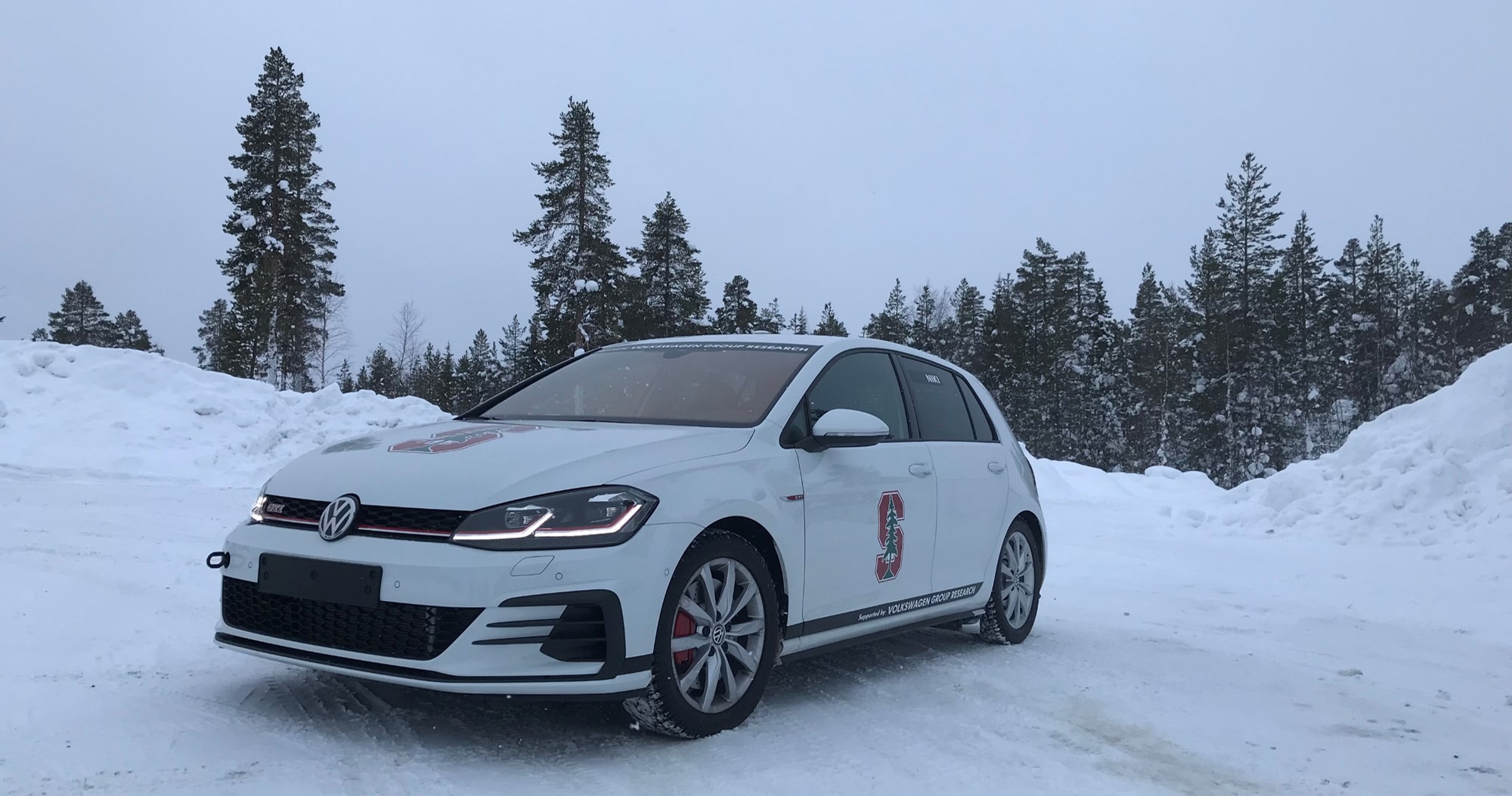}
    \caption{Volkswagen GTI Automated Test Platform}
    \label{niki}
\end{figure}

Another RMPC instantiation is Tube MPC, which reduces pessimism by solving for control policies instead of open-loop control sequences \cite{Mayne2014ModelPromise}. Closed-loop policies can react to future events and therefore pursue higher performance in the near term. Tube MPC attempts to guarantee the set of possible disturbed states is contained within problem constraints \cite{Blanchini1990ControlDisturbances}. K{\"o}gel \textit{et~al.} presented a formulation to further reduce conservatism by combining the effects of estimation error and model uncertainty \cite{Kogel2017RobustConservatism}.

Robust MPC remains a principled strategy for systems that require \textit{guaranteed} stability, like a nuclear reactor \cite{Eliasi2012RobustPlant}. However, it can be impracticably cautious when uncertainties are numerous and large, as is the case driving on legal roads. In this article, a \textit{selectively} robust model predictive controller is introduced which achieves responsible but practical conservatism in two ways: First, contingency scenarios package uncertainties that are likely to occur together. Whereas traditional RMPC assumes all uncertainties may agitate simultaneously, engineers can leverage expert knowledge or parameter covariance to focus on the most likely emergency events. Second, we separate nominal and contingency planning into separate horizons, and define a unique cost function for each. By separating nominal and emergency costs, the contingency trajectory is unburdened by objectives typically found in MPC cost functions. Whereas traditional RMPC would aim to maintain comfortable and efficient emergency maneuvers, Contingency MPC preserves only a feasible avoidance trajectory.

Stochastic MPC is another MPC technique to handle uncertainty. Control engineers seeking robustness with less pessimism favor building parameter ambiguity and stochastic uncertainty directly in their optimizations \cite{Saltik2018AnAspects}. In SMPC, uncertain quantities are characterized into statistical distributions, and control inputs are calculated to minimize \textit{expected} cost. In this way, the stochastic algorithms do not place undue focus on highly improbable scenarios. Farina \textit{et~al.} characterized the general problem statement, which requires a disturbance model to be defined either by a probability density function (e.g. the Gaussian) or some function which produces samples \cite{Farina2016StochasticReview}. This assumption, however, may not hold for some systems or applications.

For example, a well-defined distribution may not always exist for the coefficient of friction along a roadway. We may not feel comfortable assigning a probability density to predict a child's behavior. On the contrary, if ice is likely to exist somewhere on a winter day, we should drive in such a way that a surprise icy patch can be handled safely. If children are seen playing on the sidewalk, we should anticipate the possibility they run into the street and approach with an avoidance strategy -- a contingency plan.

One SMPC technique has been borrowed to formulate Contingency MPC. In Scenario SMPC, a prediction horizon tree is built which branches from stage-nodes by sampling from a parameter distribution. At each stage-node only one command is calculated, an important constraint to prevent Scenario SMPC from improper prognostication. Lucia \textit{et~al.} demonstrated this technique with a nonlinear model, and Krishnamoorthy \textit{et~al.} worked to improve the distributed systems over which Scenario SMPC can be decomposed \cite{Lucia2013Multi-stageUncertainty} \cite{Krishnamoorthy2018ImprovingAlgorithm}. Contingency MPC's prediction horizon structure resembles a simple scenario tree, with a single stage-node at $x_0$. At this stage, we also employ the single-command constraint, shown in Fig.~\ref{flowchart}. This constraint is central to the function of CMPC; the input $u^0$ is calculated to be valid for both nominal and contingency horizons. Thus the vehicle will track the desired path as closely as possible, while maintaining the ability to choose the contingency trajectory.

This article offers a new option for control engineers who seek robustness from foreseeable but uncertain events. In Section II, the general Contingency MPC problem is defined. Section III narrows the general problem to an linear convex program for use in automated vehicle (AV) control. In Section IV, we present an experimental demonstration of CMPC. First the test platform turns through a snow-covered corner; it then encounters ice performing the same maneuver. Finally, Section V explores the potential for CMPC to tackle other problems in control under uncertainty and outlines future steps to develop the framework.

\section{CMPC FORMULATION} 

\begin{figure}
	\centering
	\includegraphics[trim={7.1cm 8.4cm 7.1cm 9.65cm},clip,scale=.7]{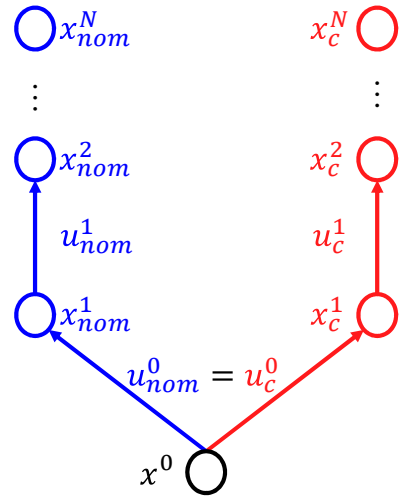} 
    \caption{Contingency MPC prediction horizon. A nominal and emergency trajectory are computed, which share a zeroth action.}
    \label{flowchart}
\end{figure}

Model Predictive Control is a receding horizon optimal control technique whose central strength is its ability to minimize future cost while explicitly respecting model and environmental constraints \cite{Richalet1978ModelProcesses}. At each time step, MPC calculates a prediction horizon (open-loop trajectory) of discrete states $x$ and inputs $u$ to optimize a cost function, subject to constraints. Upon computation, $u^0$ is deployed and a new horizon is optimized. Contingency MPC's receding horizons are illustrated in Fig.~\ref{flowchart}, where $x^0$ represents the plant's current state and two trajectories are calculated. Trajectory $x_{nom}$ is analogous to the traditional deterministic MPC horizon, containing the states we intend to guide the vehicle through by application of control inputs $u_{nom}$. On the other hand, trajectory $x_c$ represents our contingency plan of action, subject to a unique set of constraints and costs.

\setlength{\abovecaptionskip}{-8pt plus 0pt minus 0pt}
\setlength{\abovecaptionskip}{5pt plus 0pt minus 0pt}

\subsection{General Case}

The following optimization problem calculates N control inputs $\textbf{u} \in \R^{2 \cdot m}$ to guide state vectors $\textbf{x} \in \R^{2 \cdot n}$ to minimize cost function $J$ while abiding to equality constraints $H$ and inequality constraints $G$. N is the number of prediction horizon stages, \textit{n} the length of the nominal state vector, and \textit{m} the number of control inputs.

\vspace{-5 pt}

\begin{equation} \label{eq:J} \tag{1a}
J(x^{0}) =
\min_{\textbf{u}}\sum_{k=0}^{N}
j^{k}(\textbf{x}^{k},\textbf{u}^{k}) 
\end{equation}

\vspace{-5 pt}

where:

\vspace{-5 pt}

\begin{equation} \label{eq:nom_c} \tag{1b}
\textbf{x}^{k} =
	\begin{bmatrix*}[c]
    x_{nom} \\ x_{c}
 	\end{bmatrix*}^{k}
\hspace{.5cm} ; \hspace{.5cm}
\textbf{u}^{k} =
	\begin{bmatrix*}[c]
    u_{nom} \\ u_{c}
 	\end{bmatrix*}^{k}
\end{equation}

\vspace{0 pt}

such that:

\vspace{-6 pt}

\begin{equation} \label{eq:h} \tag{1c}
h(\textbf{x},\textbf{u}) = 0 \hspace{.4cm} \forall \hspace{.4cm} h \in H
\end{equation}

\vspace{-1 pt}

\begin{equation} \label{eq:g} \tag{1d}
g(\textbf{x},\textbf{u}) \leq 0 \hspace{.4cm} \forall \hspace{.4cm} g \in G
\end{equation}

\vspace{10 pt}

and for $k=0$: $\hspace{.6cm} u^0_{nom} = u^0_{c} = u^0 \hspace{2.4cm}$ (1e)

\vspace{13 pt}

Equality constraints $H$ include the dynamics model:

\vspace{-1 pt}

\begin{equation} \label{eq:dyn} \tag{1f}
\textbf{x}^{k+1} = f(\textbf{x}^{k},\textbf{u}^{k}) \hspace{.5cm} \forall \hspace{.5cm} k
\end{equation}

As in typical MPC formulations, the cost function terms $j$ can be used to encourage the nominal states $x_{nom}$ to follow a desired trajectory, maintain comfort or smoothness, minimize fuel use, etc. However, these costs need not be levied against the contingency trajectory. The separation of costs allows the contingency trajectory to focus on  maintaining a feasible emergency maneuver, unburdened by costs unimportant during crisis. Constraint sets $H$ and $G$ encode the contingency scenario. For example, constraints on $x_c$ can be added to represent an obstacle which may enter the road. Alternatively, to prepare for a dynamics-based contingency, the propagation model for $x_c^{k+1}$ can be modified from the nominal case.

The resulting optimized horizon contains two input trajectories from $x^0$: $u_{nom}$ and $u_c$. As in classical MPC, $u^0$ is deployed -- the root command for both $u_{nom}$ and $u_c$. Therefore the vehicle \textit{never chooses} between the nominal and contingency trajectories. $u_0$ tracks the desired path to the greatest extent possible, while maintaining an avoidance maneuver for the contingent event. If the contingency event occurs, CMPC is expected to return safe trajectories in subsequent time-steps. Alternatively if the contingency ceases to be relevant, the vehicle can proceed along the desired path.

\section{Linear CMPC for AV Control} 

The general CMPC formulation above can be adapted to any MPC framework. For the remainder of this article, the general form is narrowed to a convex linear optimization extended from the deterministic MPC controller developed by Brown \textit{et~al.}, and modified to solve directly for steering angle $\delta$ as presented by Zhang \textit{et~al.} \cite{Brown2017SafeVehicles} \cite{Zhang2018TireHandling}.

\subsection{Vehicle Dynamics Model}

\begin{figure}[t]
	\centering
    \includegraphics[trim={6.5cm 11.95cm 6.5cm 12cm },clip,scale=1]{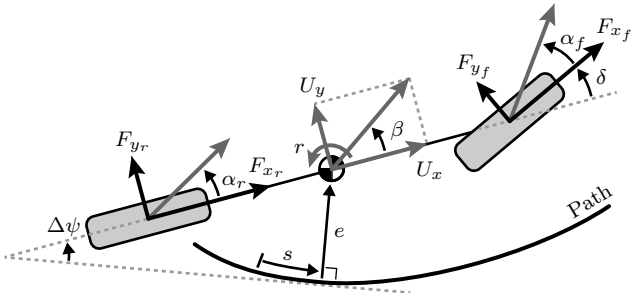} 
    \caption{Bicycle model diagram}
    \label{bicycle}
\end{figure}

To adequately handle contingency scenarios which push the limits of vehicle handling, dynamics must be modeled with an appropriate degree of fidelity. We accomplish this using a planar bicycle model with successively linearized tire forces to represent the vehicle \cite{Erlien2014IncorporatingControl}. Fig.~\ref{bicycle} illustrates this model with three position and velocity states, and forces at each axle. The position states are local to a path with curvature $\kappa$, where $s$ represents the vehicle's longitudinal progress, $e$ its lateral error from the path, and $\Delta\psi$ is the heading angle error. Velocity states are composed of longitudinal speed $U_x$, lateral speed $U_y$, and yaw rate $r$. The following differential equations govern the states' evolution:

\vspace{-5 pt}

\begin{equation} \label{eq:s} \tag{2a}
\dot{s} = U_x - U_y \Delta\psi \hspace{.6 cm}
\end{equation}
\begin{equation} \label{eq:e} \tag{2b}
\dot{e} =  U_y + U_x \Delta\psi \hspace{.6 cm}
\end{equation}
\begin{equation} \label{eq:dPsi} \tag{2c}
\dot{\Delta\psi} = r - \kappa U_x \hspace{1.53 cm}
\end{equation}
\begin{equation} \label{eq:Ux} \tag{2d}
\dot{U_x} = \frac{F_{xf} + F_{xr}}{m} + r U_y
\end{equation}
\begin{equation} \label{eq:Uy} \tag{2e}
\dot{U_y} = \frac{F_{yf} + F_{yr}}{m} - r U_x
\end{equation}
\begin{equation} \label{eq:r} \tag{2f}
\dot{r} = \frac{aF_{yf} - bF_{yr}}{I_z} \hspace{.43 cm}
\end{equation}

Forces $F_{yf}$ and $F_{yr}$ are modeled by a Fiala brush tire curve which relates lateral force to tire slip angles \cite{Pacejka2012TireDynamics}:

\vspace{-2 pt}

\begin{equation} \label{eq:af} \tag{3a}
\delta + \alpha_f = \tan^{-1}\left(\frac{U_y + ar}{U_x}\right) \hspace{.59 cm}
\end{equation}

\vspace{-5 pt}

\begin{equation} \label{eq:ar} \tag{3b}
\alpha_r = \tan^{-1}\left(\frac{U_y - br}{U_x}\right)
\end{equation}

\vspace{-2 pt}

\begin{figure}[t]
	\centering
    \vspace{-5pt}
    \setlength{\fboxrule}{0pt}
    \framebox{\parbox{3.3in}{\centering \includegraphics[trim={0cm 0cm 0cm 0.1cm},clip,scale=0.9]{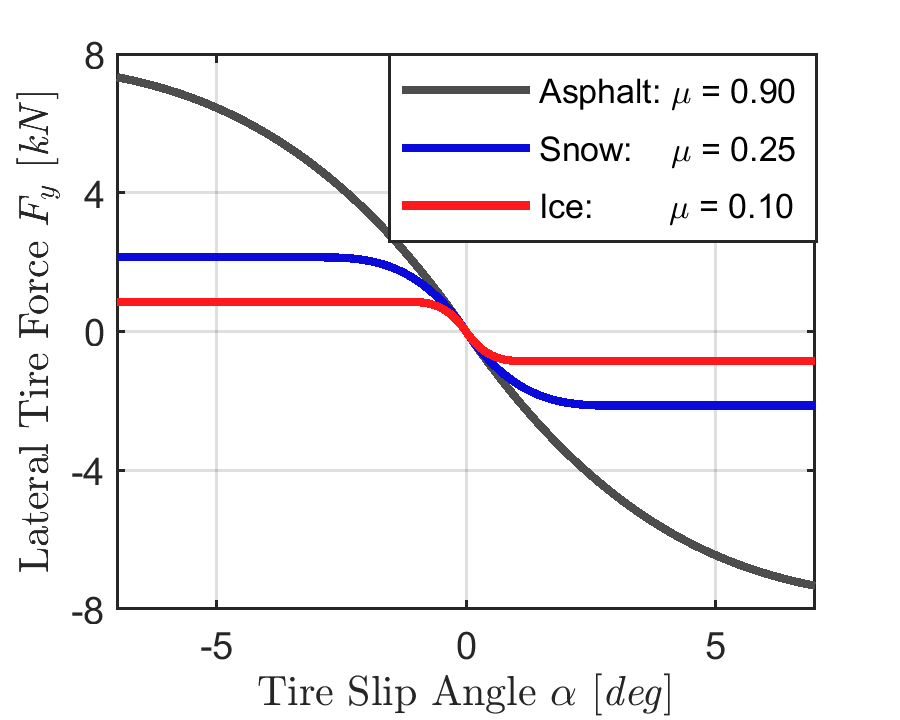}}} 
    \caption{Fiala brush tire models for asphalt, snow, and ice conditions}
    \label{fiala}
\end{figure}

\noindent In Fig.~\ref{fiala}, tire models are plotted for asphalt, snow, and ice conditions, which are the coefficients $\mu$ used in the experiments which follow.

\setlength{\abovecaptionskip}{8pt plus 0pt minus 0pt}

\vspace{2 pt}

As a first demonstration of CMPC, this formulation calculates only the steering angle $\delta$. Throttle and braking commands are calculated upstream by a feedforward-feedback scheme, as developed by Funke \textit{et~al}. \cite{Funke2017CollisionScenarios}. Including longitudinal forces into CMPC is an opportunity for future development. Thus the CMPC problem we formulate sees $F_{xf}, F_{xr}$, $U_x$, and $s$ as a known profile along the prediction horizon. The MPC state vector is then $x = [U_y \enspace r \enspace \Delta\psi \enspace e]^T$.

To complete the MPC dynamics model, the equations of motion and tire model are linearized with respect to operating points from the previous iteration's solution. The linear model was then discretized with respect to $\delta$: A zero-order hold was used for the first ten 20 ms steps, and a first-order hold was used for forty 300 ms steps to form a twelve second horizon. The number and length of time steps were chosen to capture high frequency dynamics in the short term, while extending the total horizon such that CMPC can \textit{see} far enough to prepare avoidance maneuvers. These choices were also influenced by computation time, as reported in Section IV. After discretization, the resulting affine dynamics equations take the form \cite{Zhang2018TireHandling}:

\vspace{-5 pt}

\begin{equation} \label{eq:lin_dyn} \tag{4a}
x^{k+1} = A^k x^k + B^k \delta ^k + C^k
\end{equation}

\vspace{-5 pt}

To support both nominal and contingency prediction horizons, (\ref{eq:lin_dyn}) is duplicated for both $x_{nom}$ and $x_c$ state vectors:

\vspace{-10 pt}

\begin{equation} \label{eq:lin_dyn_c} \tag{4b}
\begin{aligned}
\textbf{x}^{k+1} =
&\begin{bmatrix*}[c]
    x_{nom} \\ x_{c}
\end{bmatrix*}^{k+1} \\ =
&\begin{bmatrix*}[c]
    A_{nom} & 0 \; \\ 0 & A_{c} \;
\end{bmatrix*}^{k} \!\! \textbf{x}^{k} +
\begin{bmatrix*}[c]
    B_{nom} & 0 \; \\ 0 & B_{c} \;
\end{bmatrix*}^{k} \!\! \textbf{u} ^k +
\begin{bmatrix*}[c]
    C_{nom} \\ C_{c}
\end{bmatrix*}^{k}
\\ &\hphantom{\qquad \qquad \qquad \qquad \qquad \qquad \qquad \qquad \qquad \bigint} 
\end{aligned}
\end{equation}

\vspace{-10 pt}

\subsection{Linear CMPC Problem Statement}

The following optimization is extended from the deterministic MPC presented in \cite{Brown2017SafeVehicles} to calculate a smooth trajectory which follows a desired path while adhering to dynamics, stability, and environmental constraints.

\vspace{-5 pt}

\begin{equation} \label{eq:lin_J} \tag{5a} \begin{aligned}
\min_{\textbf{u}} \;\;
(\textbf{x}^{50})^T
\begin{bmatrix*}[c] \, 0 & 0 \\ \, 0 & Q \end{bmatrix*}
\textbf{x}^{50} + \;
\sum_{k=0}^{50} \,
(\textbf{x}^{k})^T
\begin{bmatrix*}[c] Q & 0 \, \\ 0 & 0 \, \end{bmatrix*} &
\textbf{x}^{k} \\
+ \; (\textbf{v}^{k})^T
\begin{bmatrix*}[c] R & 0 \, \\ 0 & 0 \, \end{bmatrix*}
\textbf{v}^{k} + \, 
W \begin{bmatrix*}[c] \, \sigma^{k}_{stab} \, \\ \, \sigma^{k}_{env} \, \end{bmatrix*} &
\end{aligned} \end{equation}

\vspace{-5 pt}

where:

\vspace{-10 pt}

\begin{equation} \label{eq:lin_weights} \tag{5b}
\begin{aligned}
Q =
    \begin{bmatrix*}[c]
    \, 0 & 0 & 0 & 0 \, \\
  	\, 0 & 0 & 0 & 0 \, \\
    \, 0 & 0 & 1 & 0 \, \\
    \, 0 & 0 & 0 & 1 \, \\
 	\end{bmatrix*}
\quad & ; \quad
R = 0.01 \\[5 pt]
\textbf{v}^k = 
	\begin{bmatrix*}[c]
    \, (u^k - u^{k-1})_{nom} \\
  	\, (u^k - u^{k-1})_{c\hspace{.5cm}} \\
 	\end{bmatrix*}
\quad & ; \quad
W =
	\begin{bmatrix*}[c]
  	\; 50 & 0 \\
    \; 0 & 500 \\
 	\end{bmatrix*}
\end{aligned}
\end{equation}

\vspace{10 pt}

$Q$ applies cost to heading and lateral state errors, $R$ penalizes change in steering angle (slew rate), and $W$ discourages stability and environmental constraint violation. Weights were tuned for experimental performance and chosen to establish priority among the problem's goals: Most important is driving the constraints' slack variables to zero to ensure stability and obstacle avoidance. Next, the nominal trajectory should adhere to the desired path. Lastly the nominal input trajectory should be smooth and comfortable for passengers. This minimization is subject to (\ref{eq:lin_dyn_c}) and the following constraints. First the steering angle and its slew rate are bounded:

\vspace{-5 pt}

\begin{equation} \label{eq:lin_uEnv} \tag{5c}
|\textbf{u}^k| \leq \begin{bmatrix*}[c] \delta_{max} \\ \delta_{max} \end{bmatrix*}
\quad \, ; \, \quad
|\textbf{v}^k| \leq \begin{bmatrix*}[c] v_{max} \\ v_{max} \end{bmatrix*}
\qquad \forall \quad k
\end{equation}

\vspace{3 pt}

\noindent Next, yaw rate and sideslip are confined to an invariant set, as developed by Bobier \textit{et~al.} and  Beal \textit{et~al.} \cite{Bobier2011SlidingBoundary} \cite{Beal2013ModelHandling}. This stability envelope is illustrated in Fig.~\ref{icy_stab} and enforces ${|r| \, \leq \, \mu g / U_x}$ and $|\beta| \, \leq \, \alpha_{rear,peak} + b r / U_x \,$:

\vspace{-3 pt}

\begin{equation}  \label{eq:lin_stabEnv} \tag{5d}
\begin{bmatrix*}[c] H_{nom} & 0 \; \\ 0 & H_c \; \end{bmatrix*}_{stab}
\!\!\!\!\!\! \textbf{x}^{k} \leq
\begin{bmatrix*}[c] G_{nom} \\ G_c \end{bmatrix*}_{stab}
\!\!\!\!\!\! + \sigma_{stab}
\quad \forall \quad k
\end{equation}

\vspace{5 pt}

\noindent To prevent collision with road edges, an environmental envelope encodes the inequality $e_{min} \leq e \leq e_{max}$:

\vspace{-8 pt}

\begin{equation} \label{eq:lin_envEnv} \tag{5e}
\begin{bmatrix*}[c] H_{nom} & 0 \; \\ 0 & H_c \; \end{bmatrix*}_{env}
\!\!\!\!\! \textbf{x}^{k} \leq
\begin{bmatrix*}[c] G_{nom} \\ G_c \end{bmatrix*}_{env}
\!\!\!\!\! + \sigma_{env}
\quad \forall \quad k
\end{equation}

\vspace{2 pt}

\noindent Finally the nominal and contingency trajectories are coupled via their first commands:

\vspace{-2 pt}

\begin{equation} \label{eq:lin_u0} \tag{5f}
u^0_{nom} = u^0_c
\end{equation}

\vspace{3 pt}

In the cost function and inequality constraints, block diagonal matrices are employed to weight and constrain the nominal and contingency trajectories independently. For example, the terminal cost on $\textbf{x}^{50}$ applies penalty only to the contingency trajectory's final angular and lateral deviation, an important feature to mitigate the shortcomings of a finite horizon.  No path following or input-smoothing penalty is applied to the contingency trajectory for $k < N$, however, because we only require it to obey the imposed constraints.

\section{Experimental Results and Analysis} 

\begin{figure}[b!]
	\centering
    \setlength{\fboxrule}{0pt}
    \framebox{\parbox{3.3in}{\centering \includegraphics[trim={0cm 0cm 0cm 0cm},clip,scale=1]{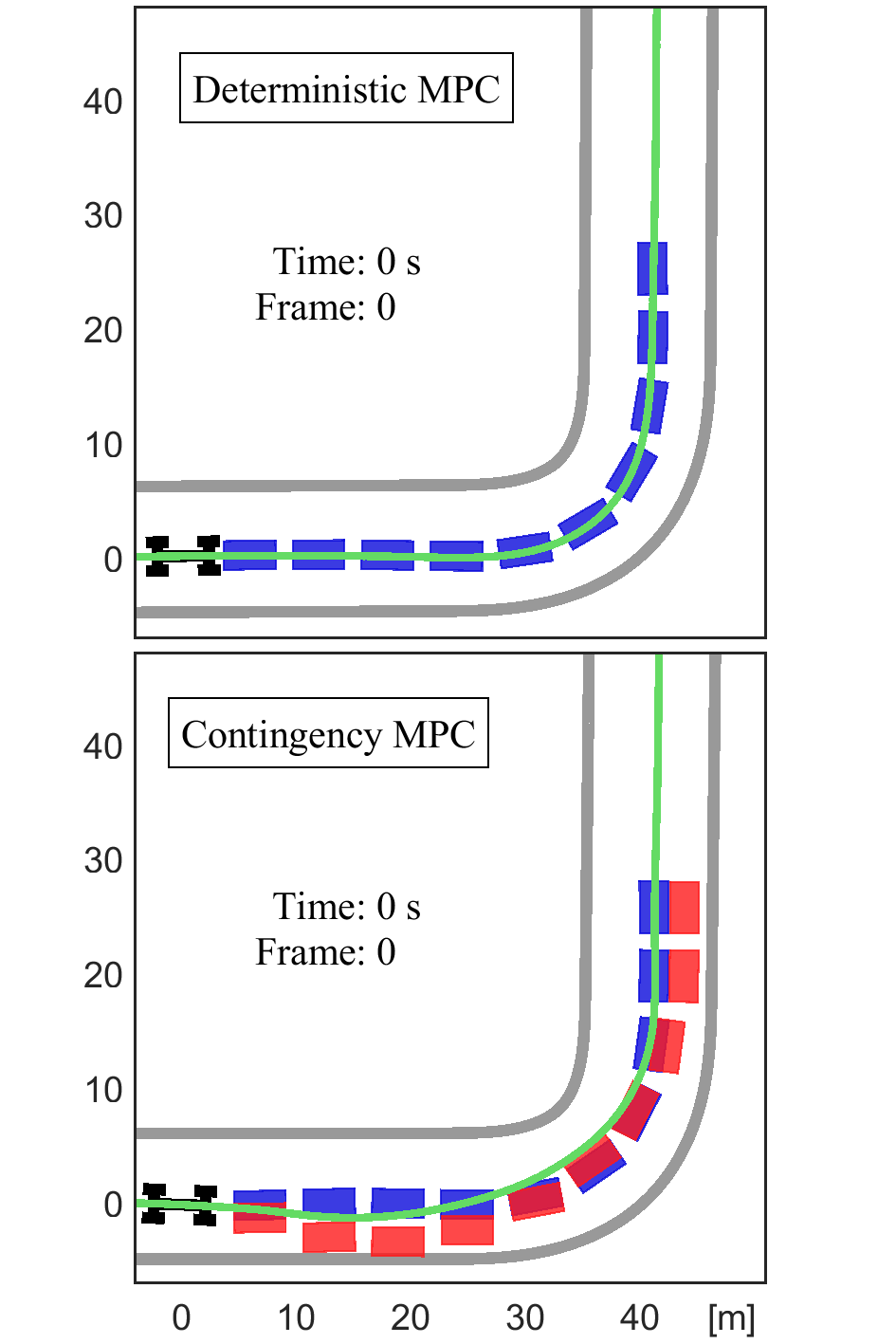}}}
    \caption{Deterministic MPC and Contingency MPC experimentally navigated a snow-covered road. Closed-loop behavior in green; nominal horizon in blue; contingency horizon in red. CMPC took a more conservative route. For clarity, Only ten of fifty horizon states are plotted from each trajectory.}
    \label{snow_exp}
\end{figure}

Contingency MPC was demonstrated in experiment using the convex linear formulation presented in Section III, deployed on an automated Volkswagen GTI (Fig.~\ref{niki}). These results compare CMPC with a deterministic MPC based on the nominal costs and constraints. In each comparison, CMPC purposefully deviated from the desired path to maintain an avoidance strategy for the anticipated contingency.


\begin{figure}[b!]
	\centering
    \includegraphics[scale=1]{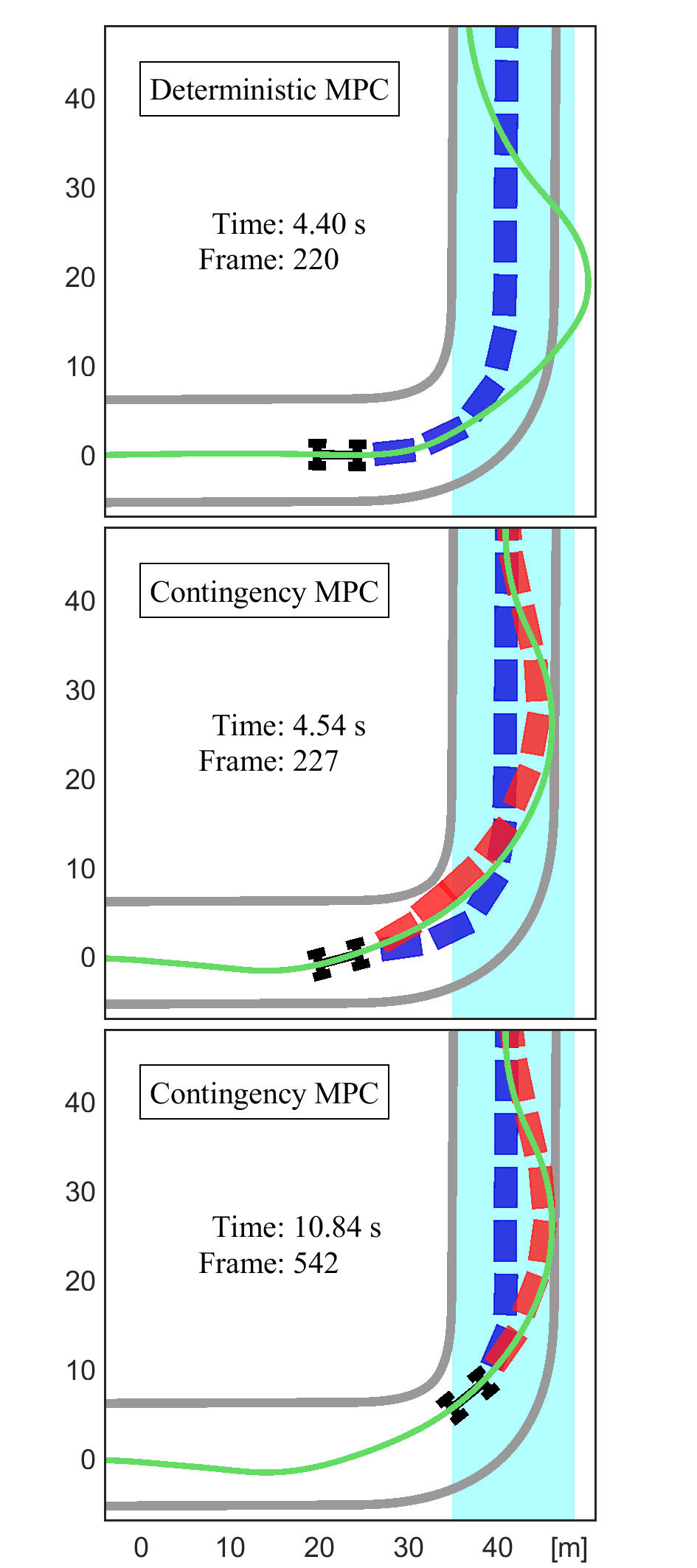}
    \caption{Deterministic MPC and Contingency MPC experimentally navigated an ice-covered left turn. DMPC did not anticipate ice and lost control. CMPC subtly deviated from the nominal path to increase robustness.}
    \label{icy_exp}
\end{figure}

The first setting is shown in Fig.~\ref{snow_exp}, a snow covered left-hand turn driven at 5 m/s -- a feasible speed for the roughly homogeneous surface. Deterministic MPC completed the maneuver while adhering closely to the desired path (the lane's centerline). The nominal dynamics modeled snow with a coefficient of friction $\mu = 0.25$.  Next, Contingency MPC completed the same maneuver while anticipating the potential for ice ($\mu = 0.10$) to cover the road. The red trajectory illustrates the contingency plan that CMPC computed for this time-step. As the vehicle proceeded under CMPC, it deviated from the desired path to keep the contingency trajectory within path boundaries. It shifted wide, then steered early to make the turn more gentle and conservative. After completing the turn, CMPC returned to the path. In this case the vehicle did not encounter ice and the minor deviation was ultimately unnecessary.

In Fig.~\ref{icy_exp}, the vehicle turned left with the same geometry onto a polished ice surface (cyan). Deterministic MPC's performance is shown in the figure's first pane. Unaware of the ice ahead, it naively followed the centerline and could not exert the lateral force necessary to track the path. It ran through the boundary (grey), and recovered after sliding beyond the ice onto snow where it regained traction. After encountering the ice, no avoidance maneuver could have ensured safety. To escape, the vehicle must have behaved differently \textit{before} the emergency.

In panes 2 and 3 of Fig.~\ref{icy_exp}, Contingency MPC demonstrated its effectiveness by repeating the Fig.~\ref{snow_exp} approach and intuitively deviating from the desired path to design a safer trajectory. The closed-loop behavior resembles the outside-inside-outside cornering technique described by professional racecar driver Taruffi, to maximize speed by minimizing curvature \cite{Taruffi2003TheRacing}. In this experiment the goal was not to race; however the vehicle's speed profile has been designed to push the scenario's limit. In order to cope with this speed (set a priori), CMPC calculated steering commands to increase the radius of curvature and decrease the tire-force necessary to stay inside the boundary. Thus, the consequence of encountering the extremely slick zamboni-polished surface was significantly mitigated by CMPC, compared to deterministic MPC. In closed-loop (green), CMPC drifted only 20 cm beyond the road edge as it exited the turn.

\begin{figure}[b!]
	\centering
    \includegraphics[scale=1]{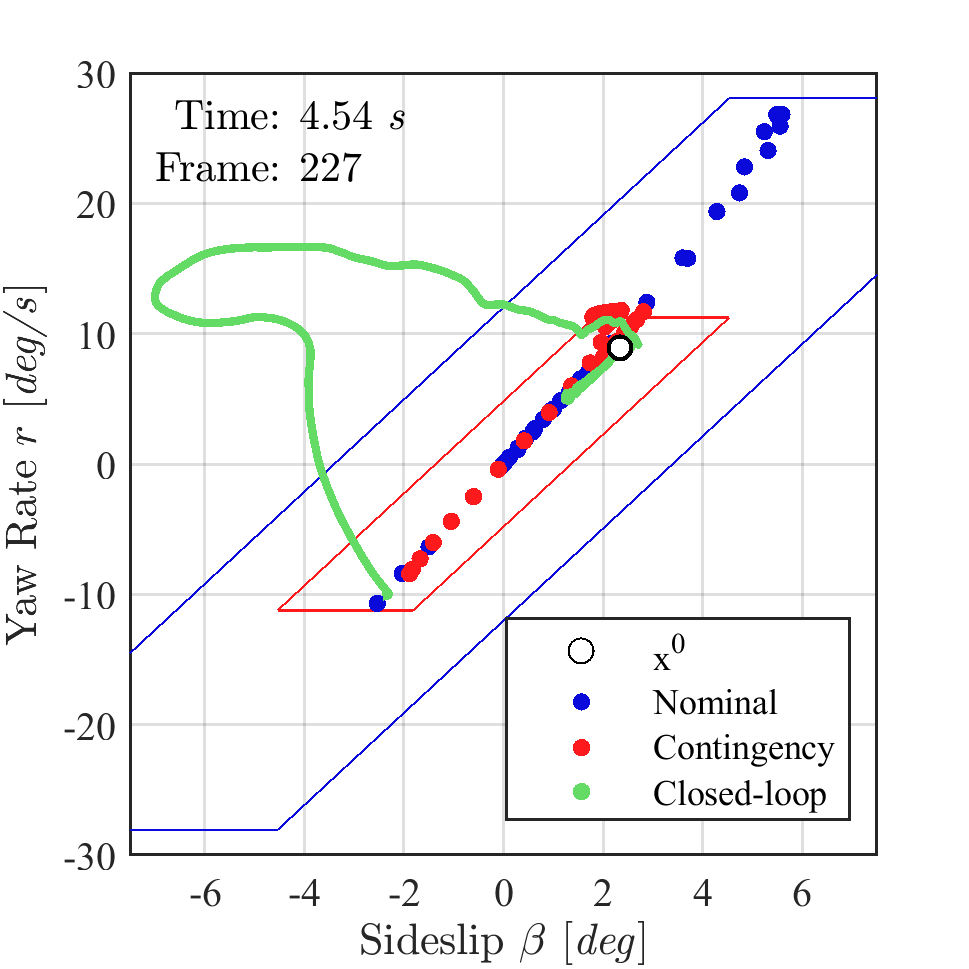}
    \caption{CMPC prediction horizons and closed-loop trajectory on the $r$~--~$\beta$ phase plane. Stability envelopes for snow (blue) and ice (red) were violated in closed-loop to prioritize staying within the road boundary.} \label{icy_stab}
\end{figure}

In Fig.~\ref{icy_stab}, velocity state data is presented from the CMPC test on ice, corresponding to the same time-step as Fig.~\ref{icy_exp}, pane 2. The parallelograms illustrate the stability envelopes from \eqref{eq:lin_stabEnv}, inside which CMPC attempted to confine the vehicle's state. Measured velocity states are plotted in green, which show the car's actual closed-loop behavior differed significantly from either plan. As it proceeded through the corner, CMPC evaluated the prospect of breaching the stability and environmental envelopes. According to the cost function weights defined in \eqref{eq:lin_weights}, CMPC allowed a greater violation of the stability guarantee as it prioritized avoidance of the road edge. Setting $\sigma_{env}$ ten times greater than $\sigma_{stab}$ influenced CMPC to trade the possibility of losing stability for a minimal collision.

The optimization for this experiment was solved using FORCES Pro on a single i7 processor \cite{Domahidi2014FORCESProfessional}. The fifty stage deterministic MPC problem converged in an average of \mbox{4.5 ms}, while Contingency MPC extended average solution times to 11 ms and up to 16 ms in some iterations. This experiment exhibited Contingency MPC's use in a dynamics-based scenario. The prospect of ice on the roadway was encoded into the contingency horizon, which caused controller to act robustly in anticipation of decreased friction.

\section{Discussion} 

The snow and ice experiments successfully demonstrated the concept and efficacy of Contingency MPC. Furthermore, the extreme polished-ice circumstance pushed this convex-linear implementation to its limit, providing insight into how CMPC and similar controllers may be improved in the future. One such limitation was the controller's focus on steering. Indeed, the simplest solution to the presented situation may have been to slow down. Future work could incorporate longitudinal and lateral inputs to CMPC by moving to a nonlinear optimization or by decoupling $F_x$ and $F_y$.

Furthermore, linearizing the low friction tire model presented difficulty and contributed to the deviation illustrated in Fig.~\ref{icy_stab}. As shown in Fig.~\ref{fiala}, the generous linear region present on asphalt nearly vanishes as friction decreases. As a result, linearizations of the low friction model lose accuracy quickly as one moves away from the operating point, allowing some CMPC iterations to predict greater force than was physically available. This problem may be abated by a strategy which maintains greater fidelity of the nonlinear model.

\section{CONCLUSION} 

In this article, Contingency MPC has been established as a credible strategy to augment a deterministic model predictive controller with robustness. In systems where potential emergency situations can be identified, CMPC maintains an avoidance trajectory while following a desired path. Experimentally, the controller successfully mitigated the effect of an abrupt decrease in road surface friction by consciously sacrificing path tracking performance to increase robustness. It leveraged the knowledge that ice may appear \textit{somewhere} in the environment, and thus planned trajectories which were prepared for its occurrence.

Several avenues appear promising for future development: Applying the framework to additional scenarios may shed valuable insight into the character and range of CMPC. For example, pedestrians may cross unexpectedly or a vehicle ahead may stop abruptly. To investigate its use within a larger system, CMPC could be integrated with an online emergency recognition program to identify contingencies in real-time. In more complex situations, several contingency events may be simultaneously probable. For these cases, CMPC could be augmented to maintain two or more contingency horizons.

\newpage

\section*{ACKNOWLEDGMENT} 

Thank you Volkswagen Group Research and Engineering Research Lab for experimental support, and Embotech GmbH for academic licensing. Alsterda is supported by Renault Group and the U.S. Dept. of Veterans Affairs G.I. Bill. Brown is supported by Renault Group and the Graduate Research Fellowship from the National Science Foundation.

\bibliographystyle{ieeetr}
\bibliography{references}

\begin{thebibliography}{10}

\bibitem{Mayne2014ModelPromise}
D.~Q. Mayne, ``{Model predictive control: recent developments and future
  promise},'' {\em Automatica}, vol.~50, no.~12, pp.~2967--2986, 2014.

\bibitem{Mayne2016RobustDirection}
D.~Q. Mayne, ``{Robust and stochastic model predictive control: are we going in
  the right direction?},'' {\em Annual Reviews in Control}, vol.~41,
  pp.~184--192, 2016.

\bibitem{Campo1987ROBUSTCONTROL}
P.~J. Campo and M.~Morari, ``{Robust model predictive control},'' in {\em Proc.
  American Control Conference}, 1987.

\bibitem{Sartipizadeh2016UncertaintyMethod}
H.~Sartipizadeh and T.~L. Vincent, ``{Uncertainty characterization for robust
  MPC using an approximate convex hull method},'' in {\em Proc. American
  Control Conference}, 2016.

\bibitem{He2014Quasi-min-maxStability}
D.~F. He, H.~Huang, and Q.~X. Chen, ``{Quasi-min-max MPC for constrained
  nonlinear systems with guaranteed input-to-state stability},'' {\em Journal
  of the Franklin Institute}, vol.~351, no.~6, pp.~3405--3423, 2014.

\bibitem{Blanchini1990ControlDisturbances}
F.~Blanchini, ``{Control synthesis for discrete time systems with control and
  state bounds in the presence of disturbances},'' {\em Journal of Optimization
  Theory and Applications}, vol.~65, no.~1, p.~29–40, 1990.

\bibitem{Kogel2017RobustConservatism}
M.~K{\"{o}}gel and R.~Findeisen, ``{Robust output feedback MPC for uncertain
  linear systems with reduced conservatism},'' in {\em Proc. 20th IFAC World
  Congress}, 2017.

\bibitem{Eliasi2012RobustPlant}
H.~Eliasi, M.~B. Menhaj, and H.~Davilu, ``{Robust nonlinear model predictive
  control for a PWR nuclear power plant},'' {\em Progress in Nuclear Energy},
  vol.~54, no.~1, pp.~177--185, 2012.

\bibitem{Saltik2018AnAspects}
M.~B. Saltik, L.~{\"{O}}zkan, J.~H. Ludlage, S.~Weiland, and P.~M. Van~den Hof,
  ``{An outlook on robust model predictive control algorithms: reflections on
  performance and computational aspects},'' {\em Journal of Process Control},
  vol.~61, pp.~77--102, 2018.

\bibitem{Farina2016StochasticReview}
M.~Farina, L.~Giulioni, and R.~Scattolini, ``{Stochastic linear model
  predictive control with chance constraints - a review},'' {\em Journal of
  Process Control}, vol.~44, pp.~53--67, 2016.

\bibitem{Lucia2013Multi-stageUncertainty}
S.~Lucia, T.~Finkler, and S.~Engell, ``{Multi-stage nonlinear model predictive
  control applied to a semi-batch polymerization reactor under uncertainty},''
  {\em Journal of Process Control}, 2013.

\bibitem{Krishnamoorthy2018ImprovingAlgorithm}
D.~Krishnamoorthy, E.~Suwartadi, B.~Foss, S.~Skogestad, and J.~Jaschke,
  ``{Improving scenario decomposition for multistage MPC using a
  sensitivity-based path-following algorithm},'' {\em IEEE Control Systems
  Letters}, vol.~2, no.~4, pp.~581 -- 586, 2018.

\bibitem{Richalet1978ModelProcesses}
J.~Richalet, A.~Rault, J.~L. Testud, and J.~Papon, ``{Model predictive
  heuristic control: applications to industrial processes},'' {\em Automatica},
  vol.~14, no.~5, pp.~413--428, 1978.

\bibitem{Brown2017SafeVehicles}
M.~Brown, J.~Funke, S.~Erlien, and J.~C. Gerdes, ``{Safe driving envelopes for
  path tracking in autonomous vehicles},'' {\em Control Engineering Practice},
  vol.~61, pp.~307--316, 2017.

\bibitem{Zhang2018TireHandling}
V.~Zhang, S.~M. Thornton, and J.~C. Gerdes, ``{Tire modeling to enable model
  predictive control of automated vehicles from standstill to the limits of
  handling},'' in {\em Proc. 14th International Symposium on Advanced Vehicle
  Control}, 2018.

\bibitem{Erlien2014IncorporatingControl}
S.~M. Erlien, J.~Funke, and J.~C. Gerdes, ``{Incorporating non-linear tire
  dynamics into a convex approach to shared steering control},'' in {\em Proc.
  American Control Conference}, 2014.

\bibitem{Pacejka2012TireDynamics}
H.~Pacejka, {\em {Tire and Vehicle Dynamics}}.
\newblock Butterworth-Heinemann, 3rd~ed., 2012.

\bibitem{Funke2017CollisionScenarios}
J.~Funke, M.~Brown, S.~M. Erlien, and J.~C. Gerdes, ``{Collision avoidance and
  stabilization for autonomous vehicles in emergency scenarios},'' {\em IEEE
  Transactions on Control Systems Technology}, vol.~25, no.~4, pp.~1204 --
  1216, 2017.

\bibitem{Bobier2011SlidingBoundary}
C.~G. Bobier, S.~Joe, and J.~C. Gerdes, ``{Sliding surface envelope control:
  keeping the vehicle within a safe state-space boundary},'' in {\em Proc.
  Dynamic Systems and Control Conference}, 2011.

\bibitem{Beal2013ModelHandling}
C.~E. Beal and J.~C. Gerdes, ``{Model predictive control for vehicle
  stabilization at the limits of handling},'' {\em IEEE Transactions on Control
  Systems Technology}, vol.~21, no.~4, pp.~1258 -- 1269, 2013.

\bibitem{Taruffi2003TheRacing}
P.~Taruffi, {\em {The Technique of Motor Racing}}.
\newblock Bentley Publishers, 2003.

\bibitem{Domahidi2014FORCESProfessional}
A.~Domahidi and J.~Jerez, ``{FORCES Professional},'' 2014.

\end{thebibliography}

\end{document}